\documentclass[aps,prl,twocolumn]{revtex4}
\usepackage{amssymb}
\usepackage{amsmath}
\usepackage{graphicx}
\usepackage{epsfig}

\begin{document}

\title{spin wave based quantum storage of electronic spin with a ring array of
nuclear spins}
\author{ Y.D. Wang$^{1}$, Y. Li $^{2}$, Z. Song$^{3}$ and C.P. Sun $^{1,3,
a,b}$ }
\affiliation{$^1$Institute of Theoretical Physics, the Chinese Academy of Science,
Beijing, 100080, China}
\affiliation{$^2$Interdisciplinary Center of Theoretical Studies, the Chinese Academy of
Sciences, Beijing, 100080, China}
\affiliation{$^3$Department of Physics, Nankai University, Tianjin 300071, China}

\begin{abstract}
We propose a solid state based protocol to implement the universal
quantum storage for electronic spin qubit. The quantum memory in
this scheme is the spin wave excitation in the ring array of
nuclei in a quantum dot. We show that the quantum information
carried by an arbitrary state of the electronic spin can be
coherently mapped onto the spin wave excitations or the magnon
states. With an appropriate external control, the stored quantum
state in quantum memory can be read out reversibly. We also
explore in detail the quantum decoherence mechanism due to the
inhomogeneous couplings between the electronic spin and the
nuclear spins.
\end{abstract}

\pacs{PACS number:05.40.-a,03.65.-w,42.25.Bs,73.23.Hk}
\maketitle

\section{Introduction}

A practical protocol of information processing has to involve the storage
and retrieval of information. In the fast-developing field of quantum
information and quantum computation, a coherent storage process is
particularly indispensable because the quantum information carried by some
qubits manipulable is extremely fragile in usual and thus needs to be stored
before decoherence happens \cite{q-inf,za-store}. Most existing schemes,
e.g., \cite{lukin1,sun-prl} concern about the quantum storage of photon
state instead of a qubit (a two-level system) state while the latter is more
necessary in the universal quantum computation. Recently, some
investigations \cite{lukin4,zoller,exp} are devoted to explore universal
quantum storage based on the solid state system since it is widely believed
that the practical quantum computation should based on the scalable solid
state system.

Among them, a novel protocol \cite{lukin4} was presented to store the
quantum information of electronic spin in the ensemble $N$ nuclei confined
in semiconductor quantum dot. Physically this storage process can be
described as a reversible map from the electronic spin state onto the
collective spin state of the surrounding nuclear ensemble. Because of the
long decoherence time of the nuclear spins -- up to seconds \cite{lukin4},
it can be regarded as a long-lived quantum memory for electronic spin qubit,
in which the information stored in them can be robustly preserved.

To analyze the universal applicability of this protocol in practice, we
found that \cite{szs}, only in the low excitation with large $N$ limit and
the relatively homogeneous couplings of electron to the nuclei, can the
many-nuclei system behave as single mode boson to serve as an efficient
quantum memory. The large number of nuclei in such quantum memory implies
that many nuclei crowd in a single quantum dot and the inter-nucleon
interaction may not be neglected in this case. This is just the first
motivation of the present paper to consider the influence of the
inter-nucleon interaction on the quantum storage process.

On the other hand, the low excitation condition requires a ground state with
all spins oriented. Usually it can only be prepared by applying a magnetic
field polarizing all spins along a single direction. However, for a free
nucleon ensemble without the external magnetic field, there is not such a
preferred ground state with spins pointing to the same direction and thus
the low excitation condition can not be satisfied automatically. But one can
recognize that a ferromagnetic spin chain -- the Heisenberg chain-- usually
has a spontaneous magnetization in association with the mechanism of
spontaneous symmetry breaking \cite{szbs}, which naturally offers a
polarization ground state. The intrinsic interaction between spins
correlates the nuclei to form magnon, the collective excitation mode of spin
wave, even without the external magnetic field. This observation also
motivates us to explore the possibility of using a ferromagnetic quantum
spin system, instead of the non-interacting nuclei ensemble, to serve as a
robust quantum memory.
\begin{figure}[h]
\includegraphics[width=7.5cm,height=3.5cm]{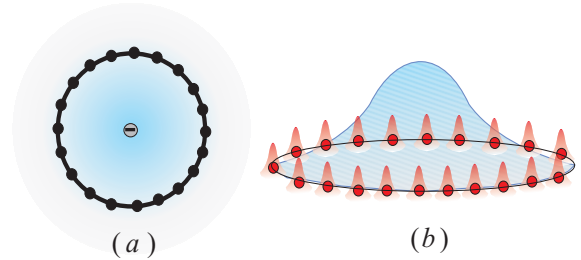}
\caption{(color online) (a) The configuration geometry of the
nuclei-electron system. The nuclei are arranged in a circle within a quantum
dot to form a ring array. To turn on the interaction one can push a single
electron towards the center of the circle along the axis that perpendicular
to the plane. (b) The overlap of symmetric wave function of electron with
those of nuclei, that results in the exchange-interactions.}
\end{figure}
With these considerations, in this paper, we present and study a protocol to
implement quantum storage of the electronic spin state to a ring-shape array
of interacting nuclei(Fig.1a). Under appropriate control of exchange
interaction between the electron spin and nuclear spins by adjusting the
overlap of their wave functions (Fig.1b), an arbitrary quantum state of the
electronic spin qubit, either pure or mixed state, can be coherently stored
in the collective mode of the nuclear spin chain, the magnon, and read out
in reverse. This paper is organized as follows: In Section II, we present
the model of quantum memory element. The quantum storage process based on
this model is illustrated in Section III and the inhomogeneity induced
decoherence effect is analyzed in detail in Section IV. In Section V,
conclusions and remarks are provided.

\section{Quantum memory unit based on the nuclear ferromagnetic spin chain
system}

To start with, we illustrate the configuration of our proposal for the
solid-state based quantum memory in Fig.1a. $N$ nuclei are arranged in a
circle within a quantum dot to form a spin ring array. A single electron is
just localized in the center of the ring array surrounded by the nuclei.
Because of the long decoherence time of the nuclei system, which is just the
motivation of this protocol, we neglect the decoherence effect due to the
damping of nuclei in the following discussions. The interaction of the
nuclear spins is assumed to exist only between the nearest neighbors while
the external magnetic field $B_{0}$ threads through the spin array.

Then the electron-nuclei system can be modelled by a Hamiltonian $%
H=H_{e}+H_{n}+H_{en}$. Here
\begin{equation}
H_{e}=-g_{e}\mu _{B}B_{0}\sigma _{z}.
\end{equation}%
corresponds to the Zeeman energy of electron with the Lande $g$ factor $g_{e}
$ and Bohr magneton $\mu _{B}$ in a magnetic field $B_{0}$ along the $z$%
-direction.

\begin{equation}
H_{n}=-g_{n}\mu _{n}B_{0}\sum_{l=1}^{N}S_{z}^{\left( l\right)
}-J\sum_{l=1}^{N}\mathbf{S}^{\left( l\right) }\mathbf{\cdot S}^{\left(
l+1\right) }.
\end{equation}%
represents the Hamiltonian for an ensemble of nuclear spins with the Lande $g
$ factor $g_{n}$, the nuclear magneton $\mu _{N}$. $\mathbf{S}^{\left(
l\right) }$ is the nuclear spin of the $l$-th site. the first term of $H_{n}$
is the Zeeman energy of the nuclear spins and the second one is the exchange
energy including dipole-dipole interaction between the nearest neighbor
nuclear spins. The coupling strength $J>0$\ and\ this spin ensemble acts as
a ferromagnetic spin chain.

\begin{equation}
H_{en}=\frac{\lambda }{2N}\left( \sigma _{+}\sum_{l=1}^{N}S_{-}^{\left(
l\right) }+h.c\right) .  \label{*}
\end{equation}%
is the hyperfine contact interaction between the $s$-state conduction
electron and the nuclei in the quantum dot. The Pauli matrices $\mathbf{%
\sigma }$ is the electronic spin. The magnitude of each coupling term $%
\propto \sigma _{+}S_{-}^{\left( l\right) }+h.c$ depends on the overlap of
the "large" wave function and the "localized " wave function of each nucleon
schematically illustrated in Fig.1 b and is estimated to be in the order of $%
10^{-1}$GHz with the same parameter in ref (\cite{lukin4}). The electronic
wave function is supposed to be cylindrical symmetric, e.g., the s-wave
component. Thus the coupling coefficient $\lambda \propto |\psi \left(
\mathbf{r}\right) |^{2}$ defined by the electron wave function $\psi \left(
\mathbf{r}\right) $ is almost homogenous for all the $N$ nuclei in the ring
array.

The denominator $N$ in Eq.(\ref{*}) originates from the envelope
normalization of the localized electron wave-function \cite%
{lukin4,zoller,exp}.

After the Holstein-Primakoff transformation
\begin{eqnarray}
S_{z}^{\left( l\right) } &=&a_{l}^{\dag }a_{l}-s,  \notag \\
S_{-}^{\left( l\right) } &=&\left( \sqrt{2s-a_{l}^{\dag }a_{l}}\right) a_{l}
\end{eqnarray}%
in terms of bosonic annihilation operator $a_{l}$, the Hamiltonian $H_{n}$
can be expressed in the low excitation limit
\begin{equation}
\left\langle a_{l}^{\dag }a_{l}\right\rangle \ll 2s,
\end{equation}%
with the bosonic excitations as%
\begin{eqnarray}
H_{n} &=&\left( 2Js-g_{n}\mu _{n}B_{0}\right) \sum_{l=1}^{N}a_{l}^{\dag
}a_{l}  \notag \\
&&-Js\sum_{l=1}^{N}\left( a_{l}^{\dag }a_{l+1}+h.c\right)   \label{***1}
\end{eqnarray}%
while the Hamiltonian of hyperfine contact interaction can be re-written as
\begin{equation}
H_{en}=\sqrt{\frac{s}{2}}\frac{\lambda }{N}\left( \sigma
_{+}\sum_{l=1}^{N}a_{l}^{\dag }+h.c\right) .  \label{***}
\end{equation}%
which is modelled as a system with one spin and $N$ interacting bosons.

The discrete Fourier transformation
\begin{equation}
a_{l}=\frac{1}{\sqrt{N}}\sum_{k=1}^{N}e^{i\frac{2\pi kl}{N}}b_{k},
\end{equation}%
is used to diagonalize the total Hamiltonian as
\begin{equation}
H=H_{N}+\sum_{k=1}^{N-1}\omega _{k}b_{k}^{\dag }b_{k}
\end{equation}%
where
\begin{eqnarray}
H_{N} &=&\omega _{N}b_{N}^{\dag }b_{N}+\frac{\Omega }{2}\sigma _{z}+  \notag
\\
&&\lambda \sqrt{\frac{s}{2N}}\left( \sigma _{+}b_{N}+\sigma _{-}b_{N}^{\dag
}\right) .  \label{***2}
\end{eqnarray}%
is the Jaynes-Cummings (JC) type Hamiltonian where $\Omega =2g^{\ast }\mu
_{B}B_{0}$. The new boson operators $b_{k}$ and $b_{k}^{\dag }$ defined by
the above equation (2) describe the the spin wave excitation, the magnon,
with the dispersion relation
\begin{equation}
\omega _{k}=g_{n}\mu _{n}B_{0}+2Js-2Js\cos \frac{2\pi k}{N}.  \label{****}
\end{equation}

The above results show that $H$ only contains the interaction of the $N$-th
magnon mode with the electronic spin while the other $N-1$ magnons are
decoupled with it. Actually this is the consequence of the cylindrical
symmetry induced by the homogeneity of couplings of the electronic spin to
the nuclei. As noted in Eq. ( and \ref{***}), the Hamiltonian is invariant
under any permutation of the nuclear spins. In the terminology of group
theory, the permutation symmetrical group $S_{n}$ is the symmetry
transformation group of the nuclei system and thus there exist $2Ns$
two-dimensional $H_{en}$-invariant subspaces $V_{Nn}$ spanned by
\begin{eqnarray}
\left\vert n,+\right\rangle &=&\left\vert n\right\rangle _{N}\otimes
\left\vert +\right\rangle ,  \notag \\
\left\vert n+1,-\right\rangle &=&\left\vert n+1\right\rangle _{N}\otimes
\left\vert -\right\rangle
\end{eqnarray}%
where
\begin{eqnarray}
\left\vert n\right\rangle _{N} &=&\frac{1}{\sqrt{n!}}\left( b_{N}^{\dag
}\right) ^{n}\left\vert G\right\rangle  \notag \\
&=&\frac{1}{\sqrt{n!N^{n}}}\left( \sum_{l=1}^{N}a_{l}^{\dag }\right)
^{n}\left\vert G\right\rangle
\end{eqnarray}%
is a symmetric state in the spatial configuration of the nuclear ensemble,
which is an symmetrical combination of all the bosonic excitation defined on
the $N$-th magnon mode ; $\left\vert +\right\rangle $ ($\left\vert
-\right\rangle $) denotes the electronic spin up\ (down) state. $%
n=0,1,\cdots ,2s-1$ is the total spin of the state $\left\vert
n\right\rangle _{N}$; $\left\vert G\right\rangle $ denotes the ground state
with all the nuclear spins oriented in the same direction perpendicular to
the spin chain surface. Under certain approximation, there also exist other $%
H_{en}$-invariant subspaces $V_{kn}$ ($k=1,2,\cdots N-1$) in the $k$-th
magnon mode similarly. Since the symmetric term $\sum_{l}a_{l}^{\dag }$ ($%
\sum_{l}a_{l}$) only belongs to the total symmetric sector of the $S_{n}$
group, it can not change the states out of the same sector of $S_{n}$.
Therefore, the $N$-th mode is singled out as the only coupling term.

\section{Quantum storage process}

The process of quantum information storage of the electronic spin can be
implemented in the invariant subspace with $N$-th magnon. Here, the
frequency $\omega _{N}=g_{n}\mu _{n}B_{0}$ of boson mode and the level
spacing $\Omega $ can be modulated by the external field $B_{0}$
simultaneously. Such spin-boson interaction forms the basis for ion-trap
based quantum computation as well \cite{ion}. With various physical systems,
people have proposed several spin-boson models based protocols for the
universal quantum information storage. For example, the nano-mechanical
resonator interacting with Josephson junction (JJ) phase qubits \cite%
{cleland} can be considered for universal quantum memory. We have analyzed
the progressive decoherence process of such spin-boson model with
nano-mechanical resonator \cite{wang}. It is just this idealized scheme that
motivates us to seek another more practical protocol based on the collective
bosonic excitation in various physical systems.

The quantum storage process is described as follows. Suppose the initial
state of the total system is prepared so that there is no excitation in the $%
N$ nuclei at all while the electron is in an arbitrary state
\begin{equation}
\rho _{e}\left( 0\right) =\sum_{n,m=\pm }\rho _{nm}\left\vert n\right\rangle
\left\langle m\right\vert
\end{equation}%
where $\left\vert +\right\rangle $ ($\left\vert -\right\rangle $) still
denotes the electronic spin up\ (down) state. The quantum information
carried by the electron is represented by the coefficients $\rho _{nm}$. The
initial state of the total system can then be written as
\begin{equation}
\rho \left( 0\right) =\rho _{b}\left( 0\right) \otimes \left\vert
0_{N}\right\rangle \left\langle 0_{N}\right\vert \otimes \rho _{e}\left(
0\right)
\end{equation}%
in terms of $\rho _{b}\left( 0\right) =\left\vert \{0\}\right\rangle
_{N-1}\left\langle \{0\}\right\vert $ where
\begin{equation*}
\left\vert n_{1},n_{2},\cdots ,n_{N-1}\right\rangle \equiv \left\vert
\left\{ n_{k}\right\} \right\rangle _{N-1}
\end{equation*}%
(for $k=1$, $2$, $\cdots $, $N-1$) denotes the Fock state of the other $N-1$
magnons. If we set $B_{0}=0$, then $\omega _{N}=0$ and the effective
Hamiltonian $H_{N}$ becomes
\begin{equation}
H_{N}=2\lambda \sqrt{\frac{s}{2N}}\left( \sigma _{+}b_{N}+\sigma
_{-}b_{N}^{\dag }\right) \equiv \lambda \hat{X}.
\end{equation}%
with $\hat{X}=2\sqrt{\frac{s}{2N}}\left( \sigma _{+}b_{N}+\sigma
_{-}b_{N}^{\dag }\right) $. Driven by such resonant JC\ Hamiltonian the time
evolution from $\rho \left( 0\right) $ at
\begin{equation}
t=t_{0}\equiv \frac{\pi }{\lambda }\sqrt{\frac{N}{2s}},
\end{equation}%
is just described as a factorized state
\begin{equation}
\rho \left( t_{0}\right) =\rho _{b}\left( 0\right) \otimes w_{F}\otimes
\left\vert -\right\rangle \left\langle -\right\vert ,  \label{***3}
\end{equation}%
where
\begin{equation}
w_{F}=\sum_{n,m=0,1}w_{nm}\left\vert n_{N}\right\rangle \left\langle
m_{N}\right\vert
\end{equation}%
is the state of the $N$-th magnon after storing with
\begin{equation}
w_{nm}=\rho _{nm}\exp (\frac{i}{2}\left( m-n\right) \pi ).
\end{equation}%
Here, to simplify our expression, we have defined
\begin{equation*}
\rho _{++}\equiv \rho _{00},\rho _{+-}\equiv \rho _{01},\rho _{-+}\equiv
\rho _{10},\rho _{--}\equiv \rho _{11}.
\end{equation*}%
The difference between $w_{F}$ and $\rho _{e}\left( 0\right) $ is only a
diagonal unitary transformation
\begin{equation}
U=\left(
\begin{array}{cc}
1 & 0 \\
0 & e^{i\frac{\pi }{2}}%
\end{array}%
\right)
\end{equation}%
\textit{is independent of the stored initial state} $\rho _{e}\left(
0\right) $ and can be cancelled by redefining the computational basis. This
means the quantum information carried by the electron has been transferred
to the nuclei.

The above calculation demonstrates that, by placing the electron near the
nuclei and switching off the magnetic field, the quantum information carried
by the electron spin can be mapped onto the $N$-th mode of the spin wave
after a time duration $t_{0}$. After the electron is removed away from the
nuclei, the $N$-th magnon will not undergo dynamic evolution and thus the
stored quantum information is well preserved. To read out the information
again, one just needs to reverse the operation with the same time duration $%
t_{0}$. It seems that the external magnetic field does not play a role in
this storage process, but it helps to prepare the initial state of nuclear
spins with all spins allied to its opposite direction. After this
preparation, the magnetic field can be turned off all through. The
ferromagnetic interaction among spins spontaneously breaks the isotropically
degenerate ground state into this special direction and prevents the system
from high excitations and perturbations and thus it serves to stabilize the
system. Because the total Hamiltonian $H$ decouples from the interactions
between the other $N-1$ magnons and the electronic spin, the above protocol
is still valid for any other initial state of $N-1$ magnons such as the
thermal state so long as the $N$-th mode is prepared in the vacuum state\cite%
{szs}.

For the experimental implementation of the above protocol, the manipulation
induced perturbations need to be discussed. As it seems from the above
illustrations, this proposal requires an instantaneous control of the
interaction between the electronic spin and the nuclear spins. However it is
known that the interaction can never be switched on and off instantaneously.
Therefore the coupling coefficient $\lambda $ should be an analytical
function of time $t\,$, that is, $\lambda \equiv \lambda \left( t\right)$.
Actually the finite rise-and-fall time of the control pulse causes deviation
from the ideal calculation in many cases. Fortunately, in our protocol,%
\textit{\ this instantaneous time control is not really an indispensable
part as it seems}. From Eq. (\ref{***2}), we can see that the free
Hamiltonians for the electronic spin and the $N$-th magnon are both zero.
The total effective Hamiltonian
\begin{equation}
H_{N}=\lambda (t)\hat{X}.
\end{equation}%
at different instant are still commutative and then the evolution operator
can be exactly
\begin{equation}
U(t)=\exp (-i\hat{X}\int_{0}^{t}\lambda \left( t^{\prime }\right) dt^{\prime
}).
\end{equation}%
unlike the usual time order integral. Therefore, the time-dependence of the
coupling term only leads to a modification of the operation time $t_{0}$. So
long as the new operation time $t_{0}^{\prime }$ satisfying
\begin{equation}
\int_{0}^{t_{0}^{\prime }}\lambda \left( t\right) dt=\frac{\pi }{\lambda }%
\sqrt{\frac{N}{2s}},
\end{equation}%
$\rho \left( t_{0}^{\prime }\right) $ is still factorized \ as the Eq.(\ref%
{***3}) to record the quantum information of the initial state exactly.

\begin{figure}[h]
\includegraphics[width=4.5cm,height=3.5cm]{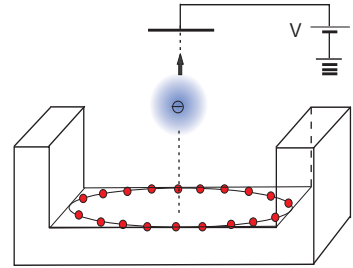}
\caption{(color online) Turn on and off of the nuclei-electron interaction
can be implemented by using an external electrical field $E$ to drive the
electron an external electrical field , which drive the electron moving
upward and downward along the axis that perpendicular to the plane.}
\end{figure}

Another difficulty in the implementation is due to the perturbation of
magnetic field caused by the electron. To store (retrieve) the information
of the electronic spin to the nuclear spin chain one needs to monitor the
coupling strength. For example we can remove the electron along the axis
perpendicular to the plane where the ring spin array locates (see Fig.2). To
maintain the symmetry of interaction between the electron and nuclear
ensemble, the axis threads through the center of the ring of spins.
Generally, the motion of electron is not at a constant velocity. The
accelerated motion generates a magnetic field according to Maxwell equation.
For example, if we use an external electrical field $E$ to drive the
electron (see Fig. 2), the acceleration of the electron is proportional to
the magnitude of $E$. The additional magnetic force will change the states
of the nuclear spin systems since both the ground state and energy spectrum
of magnons depend on the external magnetic field. This perturbation might be
fatal for the free spin chains but less serious for our correlated spin
chain since the interaction obviates the random flips of the nuclear spins.
Thus the effect of this manipulation induced perturbation is greatly
suppressed and this kind of spin chain acts as a more robust quantum memory.

\section{Decoherence induced by the inhomogeneity of couplings}

So far we have discussed the ideal case with homogeneous coupling between
the electron and the nuclei, that is, the coupling coefficients are the same
constant $\lambda $ for all the nuclear spins. However, the inhomogeneous
effect of coupling coefficients has to be taken into account if we concern
some cases beyond the strictly cylindrical symmetric effective coupling. In
this case, the quantum decoherence induced by the so-called quantum leakage
has been extensively investigated for the atomic ensemble based quantum
memory \cite{sun-you}. We now discuss the similar problems for the magnon
based quantum memory.

For the general case, $\lambda _{l}\propto |\psi \left( \mathbf{r}%
_{l}\right) |^{2}$ vary with the positions of the nuclear spins where $\psi
\left( \mathbf{r}_{l}\right) $ is the wave function of the electron at site $%
\mathbf{r}_{l}$. In this case, the Hamiltonian contains other terms besides
the interaction between the spin and $N$-th mode boson, that is, the
inhomogeneity induced interaction
\begin{equation}
V=\lambda \sqrt{\frac{s}{2N}}(\sigma _{+}\sum_{k=1}^{N-1}\chi _{k}b_{k}+h.c.)
\label{dis}
\end{equation}%
should be added in our model Hamiltonian $H$ with
\begin{equation}
\chi _{k}=\sum_{l=1}^{N}\frac{\lambda _{l}}{\lambda N}\exp [i\frac{2\pi kl}{N%
}]
\end{equation}%
are dimensionless coefficients representing the relative magnitude of
coupling.

For a Gaussian distribution of $\lambda _{l}$, e.g.
\begin{equation}
\lambda _{l}=\frac{\lambda }{\sqrt{2\pi }\sigma }\exp [\frac{-(l-1)^{2}}{%
2\sigma ^{2}}]
\end{equation}%
with width $\sigma $ and $\lambda _{1}=\lambda $, the corresponding
inhomogeneous coupling is depicted by
\begin{equation}
\chi _{k}=\frac{1}{N}\sum_{l-1=0}^{N-1}\frac{\exp [\frac{-(l-1)^{2}}{2\sigma
^{2}}+i\frac{2\pi kl}{N}]}{\sqrt{2\pi }\sigma }
\end{equation}%
Fig.3 shows the magnitude of $\chi _{k}$ for the different Gaussian
distributions of $\lambda _{l}$ with different widths $\sigma $. It can be
seen that the coupling is stronger for the modes near mode $1$ and $N-1$
than those far away from them. When the interaction gets more homogenous
(with larger $\sigma $), the coupling coefficients $\chi _{k}$ for all the $%
1 $ to $N-1$ modes become smaller. It can be imagined that when the
distribution is completely homogeneous (with $\sigma \rightarrow \infty $),
all the couplings with the $N-1$ magnon modes vanish and the Hamiltonian $H$
in Eq. (\ref{*}) is obtained.
\begin{figure}[h]
\includegraphics[width=6.5cm,height=4cm]{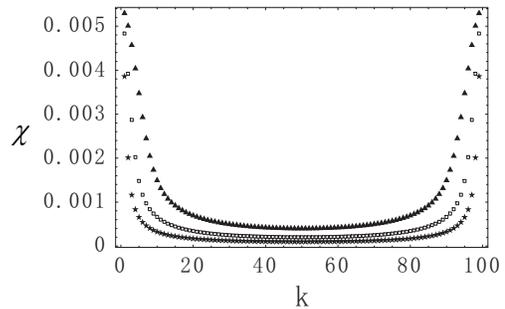}
\caption{The relative coupling coefficients $\protect\chi _{k}$ ($k=1,\cdots
,N-1$) for different Gaussian distributions. The three curves represent the
distributions with width $\protect\sigma =0.05N$ (in triangle), $\protect%
\sigma =0.1N$ (in hollow box), $\protect\sigma =0.2N$ (in pentagram)
respectively. Here $N=100$.}
\end{figure}

For the above model for quantum dissipation of two-level system \cite{legget}%
, various approaches have been presented to deal with the dynamic evolution
of the spin-boson model for different types of bath, such as the Markovian
method \cite{Louisell} and the exact solutions in the Ohmic regime \cite%
{sun-yu}. Most recently, DiVincenzo and Loss explored the calculation of the
next order Born approximation for the dissipation of such spin-boson model
\cite{loss}. In the following we will adopt a rather direct method to
analyze the decoherence problem for two limiting case.

If $N$ is so large that the spectrum of the quantum memory is
quasi-continuous, this model is similar to the "standard model" of quantum
dissipation for the vacuum induced spontaneous emission \cite{Louisell}. The
$N-1$ magnons will induce the quantum dissipation of the electronic spin
with decay rate%
\begin{equation}
\gamma =2\pi \sum_{k=1}^{N}\frac{\lambda ^{2}s|\chi _{k}|^{2}}{2N}\delta
\left( \omega _{k}-2\lambda \sqrt{\frac{s}{2N}}\right) .
\end{equation}
Let $\left\vert \Psi (t)\right\rangle $ be the ideal evolution governed by
the expected Hamiltonian without dissipation while the realistic evolution $%
\left\vert \Psi ^{\prime }(t)\right\rangle $ governed by the Hamiltonian
with dissipation term Eq. (\ref{dis}). Suppose the initial state of the
electron is
\begin{equation*}
\left\vert \Psi (0)\right\rangle =\frac{1}{\sqrt{2}}\left( \left\vert
+\right\rangle +\left\vert -\right\rangle \right) ,
\end{equation*}
we can obtain the analytical result of the fidelity
\begin{eqnarray}
&&F(t)=|\langle \Psi (t)\left\vert \Psi ^{\prime }(t)\right\rangle |=\frac{1%
}{2}+\frac{1}{2}e^{-\frac{\gamma }{2}t}\times  \notag \\
&&\sec \varphi (\cos gt\cos \left( \Delta _{1}^{\prime }t+\varphi \right)
+\sin gt\sin \Delta _{1}^{\prime }t),
\end{eqnarray}
where
\begin{eqnarray}
\varphi &=&\arcsin \sqrt{\frac{2N\gamma ^{2}}{\lambda ^{2}s}},  \notag \\
g &=&\lambda \sqrt{\frac{s}{2N}},\Delta _{1}^{\prime }=\sqrt{g^{2}-\gamma
^{2}}.
\end{eqnarray}

Fig. 4 shows the curve of the fidelity $F(t)$ varying with time $t$. We can
see that the fidelity exhibits exponential decay behavior with sinusoidal
oscillation. At the instance that the quantum storage process is just
finished, the fidelity is about $1-\pi \gamma /8g$. Therefore, the deviation
from the ideal case with homogeneous couplings is very small for $\gamma
/g<<1$. Since the ring-shape spin array with inhomogeneous coupling is just
equivalent to an arbitrary Heisenberg spin chain in the large $N$ limit, the
above arguments means that an arbitrary Heisenberg chain can be used for
quantum storage following the same strategy above if $\gamma /g$ is small,
i.e., the inhomogeneous effect is not very strong.
\begin{figure}[h]
\includegraphics[width=6.5cm,height=4cm]{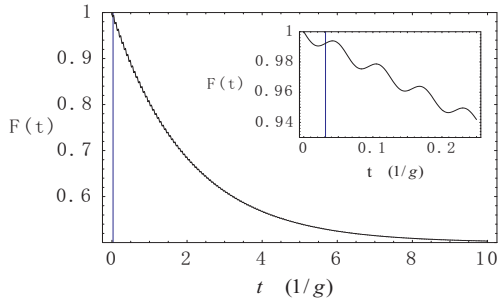}
\caption{The fidelity $F\left( t\right) $ in the large $N$ limit. The
vertical line indicates the instant\ $\protect\pi /2g$, at which the quantum
storage is just implemented. Here $\protect\gamma /g=0.02$. The inset shows
the decaying oscillation with details of $F\left( t\right) $ in a small
region near the instant $\protect\pi /2g$. Here $t$ is in the unit of $1/g$.}
\end{figure}

On the other hand, if $N$ is small, the spectrum of the quantum memory is
discrete enough to guarantee the adiabatic elimination of the $N-1$ magnon
modes, i.e.,
\begin{equation}
\lambda \sqrt{\frac{s}{2N}}\frac{\chi _{k}}{|\omega _{k}|}<<1
\end{equation}%
for the $N-1$ magnon modes. As a consequence of this adiabatic elimination,
the effective Hamiltonian
\begin{eqnarray}
H_{eff} &=&\sum_{k=1}^{N-1}\omega _{k}b_{k}^{\dag }b_{k}+H_{N}  \notag \\
&&+\sum_{k,k^{\prime }=1}^{N-1}\Omega _{kk^{\prime }}(b_{k}b_{k^{\prime
}}^{\dag }\left\vert e\right\rangle \left\langle e\right\vert -b_{k}^{\dag
}b_{k^{\prime }}\left\vert g\right\rangle \left\langle g\right\vert )
\end{eqnarray}%
can be obtained by ignoring the fast oscillating terms where
\begin{equation}
\Omega _{kk^{\prime }}=\frac{\lambda ^{2}s(\omega _{k^{\prime }}+\omega _{k})%
}{2N\omega _{k^{\prime }}\omega _{k}}
\end{equation}

It can be seen from the above Hamiltonian that, though the electronic spin
qubit and the memory mode $b_{N}$ do not exchange energy with the other $N-1$
modes, the last term in $H_{eff}$ implies that they can record the
"which-way" information carried by the electronic qubit. Then it causes
dephasing, and also leads to mixing of different magnon modes. However, if
the initial state can be prepared in vacuum state for all the magnon modes,
only the diagonal terms concerning the magnon modes remain. The deviation
from the ideal case can also be estimated by the fidelity $F\left( t\right) $
\begin{eqnarray}
F\left( t\right) &=&\frac{1}{2}|\cos \Delta _{1}t\left( \cos \Delta
_{1}^{\prime }t-i\sin \Delta _{1}^{\prime }t\cos \xi \right)  \notag \\
&&+\sin \xi \sin \Delta _{1}^{\prime }t\sin \Delta _{1}t+e^{i\frac{\Omega
^{\prime }}{2}t}|
\end{eqnarray}%
where
\begin{eqnarray}
\Omega ^{\prime } &=&-\frac{\lambda ^{2}s}{N}\sum_{k=1}^{N-1}\frac{|\chi
_{k}|^{2}}{2\omega _{k}},  \notag \\
\Delta _{1}^{\prime } &=&\sqrt{\left( \frac{\Omega ^{\prime }}{2}\right)
^{2}+\frac{\lambda ^{2}s}{2N}}
\end{eqnarray}%
and
\begin{equation}
\cos \xi =\frac{\Omega ^{\prime }}{2\Delta _{1}^{\prime }},\sin \xi =\frac{%
\lambda }{\Delta _{1}^{\prime }}\sqrt{\frac{s}{2N}}.
\end{equation}%
Fig. 5 illustrates the time evolution of this fidelity $F\left( t\right) $.
It is a fast-oscillating function. However, under the condition of adiabatic
elimination, it is a slow-varying function around the storage instance $%
t=\left( \pi /\lambda \right) \sqrt{N/2s}$ and
\begin{equation*}
F\left( t=\frac{\pi }{\lambda }\sqrt{\frac{N}{2s}}\right) \simeq \frac{1}{2}.
\end{equation*}%
\begin{figure}[h]
\includegraphics[width=6.5cm,height=4cm]{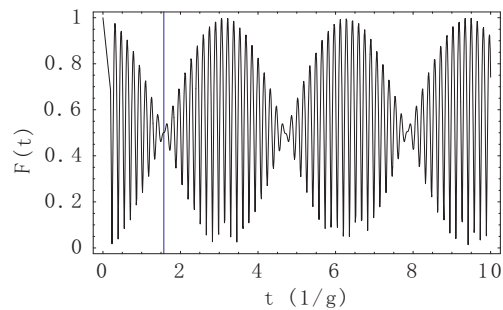}
\caption{The fidelity $F\left( t\right) $ in small $N$ limit. The vertical
line indicates the instant\ $\protect\pi /2\Delta _{1}$\ to implement
quantum storage. Here $g/2\Omega ^{\prime }$ is $0.025$ and $t$ is in the
unit of $1/g$.}
\end{figure}

Thus we can conclude that in the presence of inhomogeneity the fidelity is
still high in the large $N$ limit. Therefore, to suppress the decoherence
induced by other modes of magnon, $N$ should be made large. However, as
noted from Eq. (\ref{****}), the energy spectrum is related to $N$ and it is
quasi-continuous in the large $N$ limit. If the energy spacing is smaller
than $k_{B}T$ ($k_{B}$ is the Boltzman constant), the thermal fluctuation
will wash out the quantum coherent effect. This should be avoided by using a
spin chain with
\begin{equation}
N\leq \pi \left( \arcsin \sqrt{\frac{k_{B}T}{4Js}}\right) ^{-1}.
\end{equation}%
The value of $N$ is restricted by the two requirement and the optimum number
of $N$ can be found within this range considering practical parameters in
experiments.

\section{Conclusions and remarks}

In conclusion, we have proposed a novel protocol of universal quantum
storage for the electronic spin qubit based on the ring array of nuclear
spin. Here, the quantum memory unit is the spin wave excitation (magnon) of
the spin chain. We illustrated how the quantum information of an arbitrary
state of a two-level system can be coherently mapped from the electron to
the spin wave quantum memory.

The mechanism for quantum decoherence induced by the inhomogeneity of
couplings between the electronic spin and the nuclear spins is also
investigated. In the presence of inhomogeneity, the interaction with other
magnon modes is involved and the time evolution of the whole system
inevitably deviates from the ideal case. This results in the less-than-unity
fidelity. What' more, some other problems arise. For example, if the initial
state is a superposition of the states $\left\vert \left\{ n_{k}\right\}
\right\rangle $ with different $\sum_{k}n_{k}$, this added term leads to the
entanglement of the electronic spin and the magnon states so that the two
parts can not be factorized. Thus the quantum information storage can not be
implemented effectively. Therefore this inhomogeneity imposes a strict
restriction on the initial state of magnon modes. Another obstacle is the $z$
-component coupling between the nuclear spins and the single electron spin.
In more general cases this coupling can not be neglected. For the
homogeneous couplings, if the $N-1$ magnons are prepared in a specific
subspace spanned by $\left\vert \left\{ n_{k}\right\} \right\rangle $ that $%
\sum_{k}n_{k}$ is the same, our protocol still works efficiently even
including such $z$-component coupling. For the case with inhomogeneous
couplings, this term results in the additional interactions among the
magnons. This greatly complicates the analysis since the decoherence source
now is a set of interacting bosons instead of the conventional independent
modes.

\bigskip

\textit{This work is supported by the NSFC and the Knowledge Innovation
Program (KIP) of Chinese Academy of Sciences. It is also funded by the
National Fundamental Research Program of China with No. }2001CB309310

\end{document}